\documentclass[letter,10pt]{article}
\usepackage[margin=1in]{geometry}
\usepackage{color}
\usepackage{caption}
\usepackage{amsmath}
\usepackage{graphicx}
\usepackage{subcaption}
\usepackage{epstopdf}
\usepackage{algorithm}
\usepackage{algpseudocode}

\begin{document}

\begin{titlepage}
\title{Low-density locality-sensitive hashing boosts metagenomic binning}
\author{Yunan Luo$^{1,4}$, Jianyang Zeng$^1$, Bonnie Berger$^{2,3,*}$ and Jian Peng$^{4,*}$}
\date{}
\maketitle
\setcounter{page}{0}

{\small
\noindent{$^{1}$ Institute for Interdisciplinary Information Sciences, Tsinghua University, Beijing, China\\}
\noindent{$^{2}$ Computer Science and Artificial Intelligence Laboratory, MIT, Cambridge, MA, USA\\}
\noindent{$^{3}$ Department of Mathematics, MIT, Cambridge, MA, USA \\}
\noindent{$^{4}$ Department of Computer Science, University of Illinois at Urbana-Champaign, Urbana, IL, USA \\}
\noindent{$^{*}$ Corresponding authors: \textit{bab@mit.edu} and \textit{jianpeng@illinois.edu}\\}
}

\thispagestyle{empty}
\abstract{
Metagenomic binning is an essential task in analyzing metagenomic sequence datasets. To analyze structure or function of microbial communities from environmental samples, metagenomic sequence fragments are assigned to their taxonomic origins. Although sequence alignment algorithms, such as BWA, Bowtie or BLAST, can readily be used and usually provide high-resolution alignments and accurate binning results, the computational cost of such alignment-based methods becomes prohibitive as metagenomic datasets continue to grow. Alternative compositional-based methods, which exploit sequence composition by profiling local short $k$-mers in fragments, are often faster but less accurate than alignment-based methods. Inspired by the success of  linear error correcting codes in noisy channel communication, we introduce Opal, a fast and accurate novel compositional-based binning method. It incorporates ideas from Gallager's low-density parity-check code to design a family of compact and discriminative locality-sensitive hashing (LSH) functions that encode long-range compositional dependencies in long fragments. By incorporating the Gallager LSH functions as features in a simple linear support vector machine, we demonstrate that Opal provides fast, accurate and robust binning for datasets consisting of a large number of species, even with mutations and sequencing errors. Our binning model not only performs up to two orders of magnitude faster than BWA, an alignment-based binning method, but also achieves improved binning accuracy and robustness to sequencing errors. Opal also outperforms models built on traditional $k$-mer profiles in terms of both robustness and accuracy. Finally, we demonstrate that we can effectively use our binning model in the ``coarse search" stage of a compressive genomics pipeline to identify a much smaller candidate set of taxonomic origins for a subsequent alignment-based method to analyze, thus providing metagenomic binning with high scalability, high accuracy and high resolution.
}
\let\thefootnote\relax\footnotetext{This paper was selected for oral presentation at RECOMB 2016 and an abstract is published in the conference proceedings.}
\end{titlepage}

\section{Introduction}
Metagenomics techniques enable researchers to analyze the functional and genetic composition of microbial communities from environmental samples. Amplicon-based sequencing methods, which focus on the diversity of given marker genes, e.g. the 16S rRNA gene, provide efficient phylogenetic and functional diversity surveys of microbial communities. Due to the cost effectiveness of 16S rRNA sequencing, marker gene based analysis has frequently been used for studies involving large sample sets. With recent advances in next-generation sequencing (NGS) technologies, whole genome- or fragment-based metagenomics provides much richer information on and broader functional characterisics of microbial communities in the samples. For instance, novel hypotheses of microbial functions and potential enzymes have been identified through such metagenomic analysis \cite{review}.

During the past several years, high-throughput metagenomic sequencing has been extensively applied; however, the inherent complexity of metegenomic sequencing data poses a number of computational and statistical challenges for data analysis. Normally, DNA fragments, such as sequence reads or contigs, need to be  first assigned to their organisms of origin (also called ``binning"), since genes are typically sequenced from multiple, diverse organisms. After sequence fragments are assigned to taxonomic origins, downstream data analysis can be applied to elucidate the structure of microbial populations and assign functional annotations \cite{review}. Note that in this work, we focus on the whole-genome metagenomic DNA sequencing, instead of the marker-gene based or gene-centric methods that only analyze the protein-coding regions for which other protein search algorithms have been proposed \cite{diamond,entropy}.

Arguably the most popular metagenomic binning approaches are alignment-based methods. A sequence fragment is searched against a reference database with full genomes of organisms, and the highest scoring organism is assigned as the taxonomic origin. Although efficient sequence alignment algorithms, including BWA-MEM \cite{bwamem}, Bowtie2 \cite{bowtie2} and (mega)BLAST \cite{blast},
can be readily used for this purpose, the computational cost of alignment-based methods becomes prohibitive as the size of the sequence dataset dramatically grows, which is often the case in recent studies.

Another completely different binning approach is based on genomic sequence composition. Codon usage, oligonucleotide frequencies and GC content often are distinct in different genomes. Computational classification methods have exploited such differences to identify sequences with similar compositional features. Typically, a supervised classifier, such as a naive Bayesian classifier, a neural network or a support vector machine (SVM), is trained on a set of reference genome sequences to classify the origins of metagenomic fragments \cite{wang2007naive,patil2012phylopythias,vervier2015large,phymm}. Since the lengths of metagenomic fragments can vary from 200 to 10,000 basepairs, sequence compositional features are often designed to be within a fixed dimensionality. Short $k$-mers, contiguous nucleotide fragments with $k$ basepairs, have been shown to be both efficient and effective for metagenomic binning. For example, PhyloPythia \cite{mchardy2007accurate} uses an ensemble of SVM models trained on contiguous 6-mers and demonstrates good performance on large datasets. Its successor, PhyloPythiaS \cite{patil2012phylopythias}, further improves the binning accuracy by tweaking the SVM model and simultaneously including $k$-mers of multiple sizes ($k= 3,4,5,6$) as compositional features. Since compositional methods need to compute only the $k$-mer profiles for query sequence fragments, these methods are significantly faster than alignment-based methods on large datasets, although without providing alignment resolution and often suffering a moderate loss of the robustness.
While longer $k$-mers, which capture compositional dependency within larger contexts, could potentially lead to higher binning accuracy, they are more prone to noise and errors if used in the supervised setting. Moreover, incorporating long $k$-mers as features increases  computational cost exponentially and requires significantly larger training datasets. Note that there are existing methods use mid-size $k$-mers (e.g. $k=31$) but they are mainly used for fast indexing and nearest (or exact) search \cite{clark,space,lmat,kraken} but not in the supervised manner.

Here we overcome these bottlenecks of handling long $k$-mers for classification, enabling fast, accurate and robust metagenomic binning.
We introduce a novel compositional metagenomic binning algorithm, Opal, which efficiently encodes long $k$-mers using low-dimensional profiles. To use long $k$-mers as features in an SVM, we would need up to $O(4^k)$ dimensions, which becomes practically infeasible if $k\ge 16$. Inspired by the low-density parity-check codes (also known as Gallager codes) from coding theory \cite{ldpc,ldpc2}, we propose the use of a set of low-density locality-sensitive hash (LSH) functions \cite{lsh} to represent long $k$-mers or sequence fragments. We have two major conceptual advances in this work. First, although LSH has been previously used for fast sequence alignment and assembly \cite{butler,assembly}, to the best of our knowledge, it is the first time that the idea of LSH has been proposed for compositional-based metagenomic binning. Second, we have developed LSH functions based on the Gallager design for very long $k$-mers (e.g. $k=64$), which makes LSH practically applicable for this problem. Gallager codes were initially designed for error correction but we use it to design highly efficient LSH functions for fast binning of metagenomic fragments. We have also observed error tolerance for binning, which is partially due to the design of Gallager codes. Methodologically, starting from a Gallager design matrix with row weight $t$, we construct $m$ hash functions to encode high-order sequence compositions within a $k$-mer. In contrast to the $O(4^k)$ complexity it would take to  represent contiguous $k$-mers, our proposed Gallager LSH adaptation requires only $O(m4^t)$ time.
For very long $k$-mers, we can construct the Gallager LSH functions in a hierarchical way to further capture compositional dependencies from both local and global contexts. To evaluate the performance of the Gallager LSH method, we trained an SVM model with features generated by the Gallager LSH method. When tested on a large dataset with 50 microbial species, Opal achieved better binning accuracy than the traditional method that uses contiguous $k$-mer profiles as features \cite{vervier2015large,patil2012phylopythias,wang2007naive}. Moreover, our method is more robust to mutations and sequencing errors, compared to the method with the contiguous $k$-mer representation. We also compared Opal with BWA-MEM \cite{bwamem}, the state-of-the-art alignment-based method. Remarkably, we achieved up to two orders of magnitude improvement in binning speed on large datasets; our method is also substantially more accurate than BWA-MEM when the rate of sequencing errors is high (e.g., 10-15\%).
It is remarkable to show that a compositional binning approach can be as robust as or even more robust than alignment-based approaches,
in the presence of high sequencing errors or mutations in metagenomic sequence data. Finally, we demonstrate that it is possible to combine both compositional and alignment-based methods, by applying the compositional SVM with the Gallager LSH coding as a ``coarse-search" procedure to reduce the taxonomic space for a subsequent alignment-based BWA-MEM ``fine search", to enable both efficient and accurate metagenomic binning, with improved binning accuracy, metagenomic alignment and near 20 times speedup. Previously, a similar ``coarse search" approach has been proposed to speed up the metagenomic mapping of protein-coding sequences and speed up Diamond, an earlier state-of-the-art method, by 10 times \cite{entropy,diamond}. Note that with Opal, ``coarse search" is performed by a supervised method instead of a unsupervised clustering approach, thus potentially better encoding the dependency within the data and leading to a larger speedup.

\section{Metagenomic binning revisited}
Metagenomic sequencing techniques produce a large data sets of DNA fragments (e.g. reads or contigs) from environmental samples. To understand the microbial communities and functional structures within the samples, we need to first assign or bin these sequence fragments with the taxonomic origins from which they were derived to facilitate downstream analyses. A straightforward approach for metagenomic binning is through sequence assembly. Since the DNA fragments are sampled from chromosomes of some unknown species, we should be able to identify the original species if we can reconstruct the chromosomes from the sequence fragments. However, it is often not feasible to generate accurate assemblies from metagenomic sequence fragments, due to potential undersampled organisms, ambiguity among closely related species and the limited capability and complexity of existing assembly algorithms. To deal with large datasets, efficient and accurate metagenomic binning algorithms are thus a pressing need.

\subsection{Alignment-based methods}
Possibly the most widely used binning methods are based on sequence alignment. Metagenomic fragments are binned according to their sequence similarity to a reference database consisting of genomes with taxonomical annotations. Binning tools, including MEGAN \cite{huson2011integrative}, incorporating sequence alignment programs such as (mega)BLAST and BWA-MEM and assigning taxonomic groups or organisms, have been successfully applied in many studies. Although alignment-based methods can provide high accuracy and high resolution, the demanding requirement of computational cost makes them prohibitive for large metagenomic sequence datasets, as one must align each fragment to every genome in the reference database.

\subsection{Compositional-based methods}
Instead of the time-consuming sequence alignment, sequence compositional-based binning methods exploit the sequence characteristics of metagenomic fragments and apply machine learning classification algorithms to assign putative taxnomic origins for all fragments. Since classifiers, such as support vector machines, are trained on the whole reference genome sequences beforehand, compositional methods normally are substantially faster than alignment-based methods on large datasets. The rationale of compositional-based binning methods is based on the fact that different genomes have different conserved sequence composition patterns, such as GC content, codon usage or a particular abundance distribution of consecutive nucleotide $k$-mers. To design a good compositional-based algorithm, we need to extract informative and discriminative features from the reference genomes. Most existing methods, including PhyloPythia(S) \cite{patil2012phylopythias,mchardy2007accurate}, use the $k$-mer frequency to represent sequence fragments. In the rest of this section, we will give a brief review of compositional-based methods. Here we also want make clear that there are existing methods which utilize mid-size $k$-mers (e.g. $k=31$) for fast indexing and nearest (or exact) search but not in the supervised manner \cite{clark,space,lmat,kraken}. This work is focused on comparisons of fragment feature representation for supervised binning, so we will leave comparisons to these methods in the future work.

{\noindent \bf $K$-mer profile}. We assume that a sequence fragment $s \in \Sigma^L$, where $\Sigma = \{A,T,G,C\}$, contains $L$ nucleotides. A $k$-mer, with $k < L$, is a short word of $k$ contiguous nucleotides. We define the $k$-mer profile of $s$ in a vector representation $f_k(s) \in R^{4^k}$. If we index each $k$-mer as a binary string with length $2k$, then we have a one-to-one mapping between any $k$-mer and an integer from $0$ to $2^{2k}$. In the rest of the paper, we will not distinguish the $k$-mer string with its integer presentation $i$ for notational simplicity. Each coordinate in the $k$-mer profile $f_k(s,i)$ stores the frequency of $k$-mer $i$ in the sequence fragment $s$. For instance, for a fragment $s=AATTAT$, its $2$-mer profile $f_2(s)$ has $4$ non-zero entries: $f_2(s,AA)=1/5$, $f_2(s,TT)=1/5$, $f_2(s,AT)=2/5$ and $f_2(s,TA)=1/5$. In this way, instead of representing a $L$-nucleotide fragment in $O(4^L)$, we can use $k$-mer profile to represent it in $O(4^k)$. Many previous studies have shown that a small $k$, e.g. $k=6$,  works reasonably well in practice, although longer $k$ can improve the binning accuracy but model training becomes a serious issue because of the high dimensionality which grows exponentially in $k$. A recent study \cite{vervier2015large} has found that even with a highly tuned indexing technique, we cannot easily handle $k$-mers with $k\ge 16$ in the RAM.

{\noindent \bf Classification}. After the $k$-mer profile has been constructed, we can use supervised machine learning classification algorithms, such as logistic regression, naive Bayes classifier and support vector machines, to train a binning model. The training data can be generated by sampling $L$-nucleotide fragments from the reference genomes with taxonomic annotations. Since metagenomic fragment can have different lengths depending on the applied sequencing technologies, it is possible to construct a number of binning models, each corresponding to a particular fragment length. Because the binning classifier often only involves vector multiplication, the speed of compositional-based binning algorithms is much faster than that of alignment-based methods, thus more suitable for large datasets. On the other hand, due to the fact that $k$-mer profile can only capture the local patterns within a fragment, existing compositional binning algorithms usually have lower binning accuracy than the alignment-based methods which compare fragments and references in a global way. In addition, compositional-based classification methods are generally more sensitive to mutations or sequencing errors, partially due to the way $k$-mer profile is constructed.

\section{Opal: Gallager locality-sensitive hashing for fragment binning}
In this work, we introduce Opal, a novel compositional-based metagenomic binning algorithm, that robustly represents long $k$-mers (e.g. $k=64$) in a compact way to better capture the long-range compositional dependency in a fragment. The key idea of our algorithm is built on locality-sensitive hashing, a dimensionality reduction technique that hashes input high-dimensional data into low-dimensional buckets, with the goal to maximize the probability of collisions for similar input data. LSH has been widely used in bioinformatics for fast indexing for sequence alignment and assembly \cite{butler,assembly}. To the best of our knowledge, it is the first time that LSH functions have been applied for compositional-based metagenomic binning. We propose to use them for represent metagenomic fragments compactly and subsequently for machine learning classification algorithms to train metagenomic binning models. Since metagenomic fragments can be very long, sometimes from hundreds of bp to tens of thousands of bp, we hope to construct compositional profiles to encode long-range dependency in long $k$-mers. To handle large $k$, we develop string LSH functions to compactly encode the global dependency with $k$-mers in a low-dimensional feature vector, as oppose to directly using a $4^k$-length $k$-mer profile vector. Although LSH functions are usually constructed in a uniformly random way, we propose a new and efficient design of LSH functions based on the idea of the low-density parity-check (LDPC) code invented by Robert G Gallager for noisy message transmission \cite{ldpc,ldpc2}. A key observation is that Gallager's LDPC design not only leads to a family of LSH functions but also makes them efficient such that even a small number of random LSH functions can well encode the long fragments. Different from uniformly random LSH functions, the Gallager LSH functions are constructed structurally and hierarchically to ensure the compactness of the feature representation and the robustness when sequencing noise appears in the data.

\subsection{Locality-sensitive hashing}
LSH is a family of hash functions that have the property that two similar objects are mapped to the same hash value \cite{lsh}. For the metagenomic binning problem, we are only interested in strings of length $k$. Then a family of LSH functions can be defined as functions $h: \Sigma^k \rightarrow R^d$ which map $k$-mers into a $d$-dimensional Euclidean space. Assume that we consider Hamming distances between $k$-mers, if we choose $h$ randomly and for two $k$-mers $s_1$ and $s_2$ with at most $r$ different positions, $h(s_1)=h(s_2)$ holds with probability at least $P_1$. For two $k$-mers $s_3$ and $s_4$ with more than $R$ different positions, $h(s_3)\neq h(s_4)$ holds with probability at least $P_2$. With the construction of a LSH family, we can amplify $P_1$ or $P_2$ by sampling multiple hash functions from the family. Compared with the straightforward $k$-mer indexing representation, the LSH scheme can be more compact and more robust. For example, we can construct LSH functions such that $d \ll 4^k$. Moreover, when a small number of sequencing errors or mutations appear in the $k$-mer, LSH can still map the noisy $k$-mer into a feature representation that is very similar to original $k$-mer. This observation is highly significant since mutations or sequencing errors are generally inevitable in the data, and we hope to develop compositional-based methods less sensitive to such noises.

One way to construct LSH functions on strings under Hamming distance is to construct index functions by uniformly sampling a subset of positions from the $k$-mer. Specifically, given a string $s$ of length $k$ over $\Sigma$, we choose $t$ indices $i_1,\ldots,i_t$ uniformly at random from $\{1,\ldots,k\}$ without replacement. Then, the spaced $(k,t)$-mer can be generated according to $s$ and these indices. More formally, we can define a random hash function $h:\Sigma^{k}\rightarrow\Sigma^t$ to generate a spaced $(k,t)$-mer explicitly:
\begin{equation}
h(s) = \langle s[i_1],s[i_2],\ldots, s[i_t]\rangle.
\end{equation}
The hash value $h(s)$ can also be seen as a $4^t$ dimensional binary vector with only the string $h(s)$'s corresponding coordinate set to $1$ and otherwise $0$. It is not hard to see that such LSH function $h$ has the property that it maps two similar $k$-mers to the same hash value with high probability. For example, consider two similar $k$-mers $s_1$ and $s_2$ that differ by at most $r$ nucleotides, then the probability that they are mapped to the same value is given by
\begin{equation}
\Pr[h(s_1)=h(s_2)] \ge \binom{k-r}{t}  \bigg/ \binom{k}{t}
\end{equation}
For two $k$-mers $s_3$ and $s_4$ that differ at least $R$ nucleotides, the probability that they are mapped to  different value is given by
\begin{equation}
\Pr[h(s_3)\neq h(s_4)] \ge 1 -\sum_{j\ge R} \binom{k-j}{t}  \bigg/ \binom{k}{t}
\end{equation}
With the family of LSH functions, we randomly sample a set of $m$ LSH functions and concatenate them together as the feature vector for a long $k$-mer. Note that the complexity of the LSH-based feature vector is only $O(m4^t)$, much smaller compared to $O(4^k)$ that is the complexity of the complete $k$-mer profile. More importantly, the LSH-based feature vector is not sensitive to errors or mutations in the $k$-mer if $m$ and $t$ are well chosen, but for the traditional $k$-mer profile, even one nucleotide change can change the feature vector completely. To compute the feature vector for a metagenomic fragment with length $L$, we first extract all $k$-mers by sliding a window of length $k$ over the sequence, and then apply $h$ on each $k$-mer to generate LSH-based feature vectors and then normalize the sum of the feature vectors by $L-k+1$. In this way, one can easily show that similar fragments can also be mapped to similar LSH-based feature vectors. After the feature vectors are generated for fragments with taxonomic annotations, we train a linear classifier for metagenomic binning. It is also fairly straightforward to show that similar fragments have similar classification responses if the coefficients of the linear classification function are bounded. One may expect that the complexity of linear classification with $k$-mer profiles would be lower since there are at most $L-k+1$ different $k$-mers in a fragment and can be computed easily using sparse vector multiplications, but we find that the LSH-based feature vector is also sparse in practice and the indexing overhead is much smaller when constructing the feature vectors, since the LSH-based method can have much smaller dimensionality. In practice, the LSH-based methods can sometimes be even faster if $m$ and $t$ are not too large.

\subsection{Gallager low-density locality-sensitive hashing}
Despite that the random LSH function family described above has a lot of nice theoretical properties, uniformly sampled LSH functions are usually not optimal in practice. Theoretical properties of LSH functions hold probabilistically, which means that we need to sample a large number of random LSH functions to make sure the bounds are tight. However, practically, we simply cannot use a very large number of random LSH functions to build feature vectors for metagenomic fragments, given  the limited computational resources. Thus it would be ideal if we could construct a small number of random LSH functions that are sufficiently discriminative and informative to represent long $k$-mers. Here we take inspiration from the Gallager code or low-density parity-check code that has been widely used for noisy communication. The idea behind the Gallager code is  similar to our LSH family but with a different purpose, namely error correction. The goal of the LDPC code is to generate a small number of extra bits when transmitting a binary string via a noisy channel \cite{ldpc,ldpc2}. These extra bits are constructed to capture the long-range dependency in the binary string before the transmission. After the message string and these extra bits have been received, a decoder can perform error correction by performing probabilistic inference to compare the differences between the message string and these code bits to infer the correct message string. In the same spirit, we here adopt the idea behind the design of the LDPC code to construct a compact set of LSH functions for metagenomic binning.

\begin{figure}[h]
      \centering
        \includegraphics[width=0.70\textwidth]{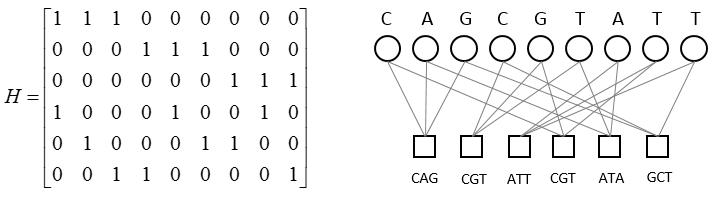}\\
    \caption{\textbf{An illustration of Gallager LSH method.} Left: an example of Gallager LDPC matrix $H$. Right: The bipartite graph corresponding to $H$. Each cycle node corresponds to a position in a $k$-mer, and each square node corresponds to a row in $H$, which generates a spaced $(k,t)$-mer.}
    \label{fig:bipartite}
\end{figure}

To construct compact LSH functions, we hope to not waste coding capacity on any particular position in the $k$-mer. While, under expectation, uniformly sampled spaced $(k,t)$-mers on average cover each position equally, with a small number of random LSH functions, it is likely that we will see imbalanced coverage among positions since the probability of a position being chosen is binomially distributed. The Gallager's design of LDPC, on the other hand, generates a subset of positions not uniformly random but make sure to equally cover each position \cite{ldpc}. So we can use the Gallager's design to generate spaced $(k,t)$-mers. The Gallager's LDPC matrix $H$ is a binary matrix with dimension $m\times k$, and has exactly $t$ 1's in each rows and $w$ 1's in each column. The matrix $H$ can be divided into $w$ blocks with $m/w$ rows in each block. We first define the first block of rows as an $(m/w)\times k$ matrix $Q$:
$$
Q=\left[
  \begin{array}{ccccccccccccccccccc}
    1 & 1 & 1 & \cdots & 1 & 1 &   &   &   &        &   &   &       &   &   &   &        &   &   \\
      &   &   &        &   &   & 1 & 1 & 1 & \cdots & 1 & 1 &       &   &   &   &        &   &  \\
      &   &   &        &   &   &   &   &   &        &   &   &\vdots &   &   &   &        &   &  \\
      &   &   &        &   &   &   &   &   &        &   &   &       & 1 & 1 & 1 & \cdots & 1 & 1\\
  \end{array}
\right],
$$
where each row of matrix $Q$ has exactly $t$ consecutive 1's from left to right across the columns. Every other block of rows is a random column permutation of the first set, and the LDPC matrix $H$ is given by:
$$
H=\left[
  %\begin{array}{c}
    Q;
    QP_1;
    \hdots;
    QP_{w-1}
  %\end{array}
\right]^T,
$$
where $P_i$ is a uniform random $n\times n$ permutation matrix for $i=1,\ldots,w-1$. An example with $k=9, t=3, m=6, w=2$ is shown in Figure \ref{fig:bipartite}.
An equivalent bipartite graph with the Gallager design matrix as the adjacency matrix also is shown.
The algorithm for constructing the LDPC design matrix is shown in Algorithm S1 in Supplementary Information.

%$$
%H=\left[
%  \begin{array}{cccccccccccc}
%    1 & 1 & 1 & 1 & 0 & 0 & 0 & 0 & 0 & 0 & 0 & 0 \\
%    0 & 0 & 0 & 0 & 1 & 1 & 1 & 1 & 0 & 0 & 0 & 0 \\
%    0 & 0 & 0 & 0 & 0 & 0 & 0 & 0 & 1 & 1 & 1 & 1 \\\hline
%    1 & 0 & 1 & 0 & 0 & 1 & 0 & 0 & 0 & 1 & 0 & 0 \\
%    0 & 1 & 0 & 0 & 0 & 0 & 1 & 1 & 0 & 0 & 0 & 1 \\
%    0 & 0 & 0 & 1 & 1 & 0 & 0 & 0 & 1 & 0 & 1 & 0 \\\hline
%    1 & 0 & 0 & 1 & 0 & 0 & 1 & 0 & 0 & 1 & 0 & 0 \\
%    0 & 1 & 0 & 0 & 0 & 1 & 0 & 1 & 0 & 0 & 1 & 0 \\
%    0 & 0 & 1 & 0 & 1 & 0 & 0 & 0 & 1 & 0 & 0 & 1 \\
%  \end{array}
%\right]
%$$

%\begin{algorithm}
%    \caption{Gallager's LDPC Matrix}
%    \begin{algorithmic}[1]
%        \State \textbf{Input:} $k$, $t$, $m$
%        \State $Q \leftarrow $ all zero $(m/w)\times t$ matrix
%        \For{$i \leftarrow 1$ to $m/w$}
%            \For{$j \leftarrow (i-1)\times t +1$ to $i\times t$}
%                \State $Q[i,j]\leftarrow1$
%            \EndFor
%        \EndFor
%        \State choose $w-1$ uniform random $n\times n$ permutation matrix $P_i$, for $i=1,\ldots,w-1$.
%        \State $H=[Q; QP_1; \ldots; QP_{w-1}]^T$.
%        \State \textbf{Output:} Gallager's LDPC Matrix $H$
%    \end{algorithmic}
%    \label{alg:gallager}
%\end{algorithm}

We use each row of $H$ to extract a spaced $(k,t)$-mer to construct an LSH function. Note that the first set of $H$ gives contiguous $t$-mers. With $m$ Gallager LSH functions, we can see that each position in a $k$-mer is equally covered $w$ times, while the same $m$ uniformly sampled LSH function is very likely to have very imbalanced coverage times for different positions because of the high variance ($=m\frac{t(k-t)}{k^2}$). {To further improve the efficiency, we construct random LSH functions with minimal overlap using a modified Gallager design algorithm. The idea is to avoid the ``4-cycles" in the bipartite graph representation, as we hope not to encode two positions together in two ``redundant" LSH functions \cite{ldpc2}. An algorithm which finds ``4-cycles" and removes them is shown in Algorithm S2 in Supplementary Information. For very long $k$-mers, we can use a hierarchical approach to generate low-dimensional LSH functions for very long-range compositional dependency in $k$-mers. We first generate a number of intermediate spaced $(k,\ell)$-mers using the Gallager's design matrix. Then from these $(k,\ell)$-mers, we again apply the Gallager's design to generate $(\ell,t)$-mers to construct the $(k,\ell,t)$ hierarchical LSH functions. Moreover, it is not hard to see that the hash functions generated from the Gallager design are also a family of LSH functions. Finally, in this work, we use a simple linear SVM to train a classification model with the Gallager LSH features. We will also test other sophisticated classifiers, such as the structured classifier that considers the taxonomic hierarchy during training, in the future. Evaluating on a small dataset with $20$ species, we found that LSH functions generated by hierarchical Gallager design is more robust and more accurate than the uniformly sampled LSH functions, indicating its efficiency for long $k$-mers (see Supplementary Figure S2).

%\begin{figure}[h]
%  \centering
%  % Requires \usepackage{graphicx}
%  \begin{subfigure}{0.49\textwidth}
%                \includegraphics[width=\textwidth]{fig/cover32.png}
%  \end{subfigure}
%\begin{subfigure}{0.49\textwidth}
%                \includegraphics[width=\textwidth]{fig/cover128.png}
%        \end{subfigure}
%  \caption{Comparison of position coverage between LSH and Gallager LSH. The left figure shows how many times a position being covered when spaced $(64,8)$-mers are generated by $m$ uniformly sampled LSH and Gallager LSH functions, respectively. The right figure compares the relationship between the number of hash functions and the classification performance.}
%  \label{fig:cover}
%\end{figure}

%\begin{algorithm}
%    \caption{Removing 4-Cycle}
%    \begin{algorithmic}[1]
%        \State \textbf{Input:} Gallager's LDPC Matrix $H$
%        \Repeat
%            \For{$i \leftarrow 1$ to $k-1$}
%                \For{$j \leftarrow i+1$ to $k$}
%                    \If{$|H[:,i]\cup H[:,j]| \ge 2$}\Comment{check if 4-cycle exists}
%                        \State $ridx \leftarrow $ row index of the first same element in $H[:,i]$ and $H[:,j]$.
%                        \State $b \leftarrow \lceil ridx / (m/w)\rceil$
%                        \State swap the elements of $H[:,i]$ and $H[:,j]$ that belong to the $b$-th block.
%                    \EndIf
%                \EndFor
%            \EndFor
%        \Until{No 4-cycle}
%        \State \textbf{Output:} 4-Cycle-free Gallager's LDPC Matrix $H$
%    \end{algorithmic}
%    \label{alg:cycle}
%\end{algorithm}

\subsection{Compositional-based binning as ``coarse search" in compressive metagenomics}
After we train the Gallager LSH-based binning classifier, we can use it as a ``coarse search" procedure in the compressive genomics manner to reduce the search space of alignment-based methods \cite{entropy}. For example, if we want to perform binning against $100$ reference organisms, instead of comparing a fragment to all reference genomes, we first apply the compositional-based binning classifier to identify a very small subset or group of putative taxonomic origins that are ranked very highly by the classifier. Then we perform sequence alignment between the fragment and the reference genomes of the top-ranked organisms. This natural combination of compositional-based and alignment-based methods provides metagenomic binning with high scalability, high accuracy and high-resolution alignments.

\section{Results}
\subsection{Experimental setting}
We downloaded 50 complete bacterial genomes from NCBI database as suggested by \cite{vervier2015large} (see Supplementary Table S1), and simulated metagenomic samples by generating fragments from these reference genome sequences. We set the coverage $c = 0.1$, and generate fragments of length $L=200$bp and $L=400$bp from the reference genome sequences. The number of fragments sampled from a genome sequence of a species, $N$, is determined by a coverage number $c$, which is defined by $c=N\times L / l$, where $l$ is then length of the whole genome sequence of the species. This coverage value is chosen since we found that larger $c$ will not further improve the performance of the classifier but it may vary if we include more species in the dataset. In addition, with a very large $c$, it would be difficult to design fair experiments, because there would be a lot substantially overlapped fragments in training and test sets. For $L=200$, we sampled 71,259 training fragments and for $L=400$, we sampled 35,631 training fragments. To assess the robustness of our method, we also simulated mutation and sequencing error in a genome sequence with rates in \{5\%, 10\%, 15\%\}. In each mutated position, the original nucleotide is substituted by one of the other three type of nucleotides with equal probability. We have also experimented on noisy data with 1\% indels and observed similar results. In this work, we use this setting to show the effectiveness of the Gallager LSH method as a proof-of-concept. We plan to further investigate other values of parameters in the future, for example, fragments with $1k$ or $10k$ nucleotides, other noise models and a much larger set of reference genomes.

To evaluate the performance of our method, we simulated a test set of fragments that are not used and with less than 50\% overlapped positions with any training fragment. The ratio between the size of training data and that of test data is 5:1. For each of the $k$-mer profile method and different Opal settings, we trained a multiclass SVM with a inner-loop cross-validation on training data only for hyperparameter selection. Here we selected SVM because it works better than several other classification methods, including naive Bayes classifier and logistic regression in our local test. Then we measured the binning performance by first computing the portion of misclassified fragments within each species, and taking the mean error rate across all species. This evaluation indicator is less biased when there are over-represented species or species with genomes of extremely imbalanced sizes in the data.

We compared the classification errors of $k$-mer profile method, alignment-based method, uniformly random LSH-based method and our Opal method based on the Gallager LSH. For the $k$-mer profile method, we constructed the 12-mer profile, which is an optimal $k$-mer profile that can be loaded into memory, as shown in a previous work \cite{vervier2015large} and also in our in-house experiment (see Supplementary Figure S1). For the alignment-based method, we chose BWA-MEM with default settings as suggested in \cite{vervier2015large}. In a in-house experiment, we also find BWA-MEM outperforms megaBLAST in terms of both speed and accuracy on our dataset. For the uniformly random LSH-based method, we sampled a set of spaced $(64,8)$-mers to construct LSH functions, denoted as LSH(64,8). For Opal, we randomly generated 32 Gallager LSH functions, denoted as Opal(64,8). We also constructed hierarchical Gallager LSH functions with the first layer generated by the Gallager design of $(64,32)$ and the second layer with $(32,8)$, denoted as Opal(64,32,8). For all LSH-based methods, we randomly drawn $32$ hash functions to construct the compositional feature vectors. Note that the dimensionality of LSH-based feature vector under this setting is $32\cdot 4^8$, much smaller than $4^{12}$, the dimensionality of $k$-mer profile. It is expected that we see better performance if more hashing functions are sampled. For Opal, we used the hierarchical (64,32,8) as the default setting.

\subsection{Comparison to previous compositional-based methods}

\begin{figure}[h]
      \centering
        \includegraphics[width=0.80\textwidth]{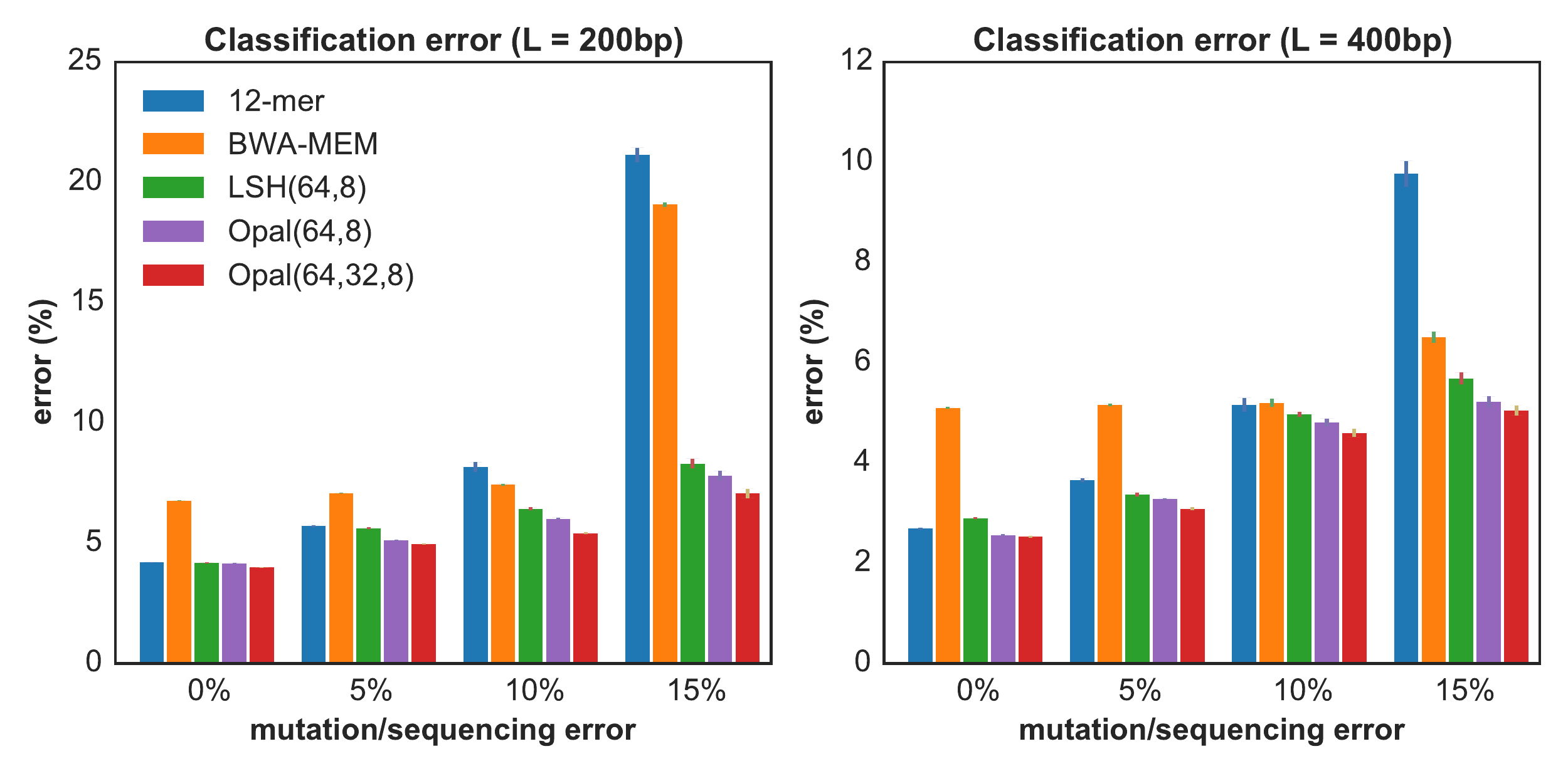}\\
    \caption{\textbf{Performance comparison of metagenomic binning methods.} }
    \label{fig:result}
\end{figure}

{\noindent \bf Opal and LSH-based methods outperforms traditional $k$-mer profile.} Due to the limit of computational resources, the $k$-mer profile method fails to handle very large $k$, while the LSH-based method can capture long range information in a very long $k$-mer. We compared the performance between $k$-mer profile and LSH-based methods. The results are shown in Figure \ref{fig:result}. Even with a smaller dimensionality, LSH-based methods, including LSH(64,8), Opal(64,8), Opal(64,32,8), achieve almost the same performance at mutation/sequencing error rate 0. As the mutation/sequencing error rate increases, the performance of the $k$-mer profile method drops dramatically. The $k$-mer profile method suffers severely from the mutations/sequencing errors and its binning error increases to 21.14\% and 9.77\% with mutation/sequencing error rate 15\%, for $L=200$ and $L=400$, respectively. The uniformly random LSH-based method, however, shows the robustness to the mutations/sequencing errors. Its misclassification rate increases slightly, with only around 5\% for $L=200$ and 3\% for $L=400$, while the misclassification rate of the $k$-mer profile method increases by 20\% and 7\%, respectively.

{\noindent \bf The Gallager LSH is more efficient than the uniformly random LSH.} Next, we showed the comparison between the Gallager low-density LSH and the uniformly random LSH. The Gallager LSH has an advantage over random LSH in that each position in the contiguous $k$-mer gets equal coverage when the spaced $(k,t)$-mers are generated, hence more efficient when the number of hash functions sampled is practically manageable (see Supplementary Figure S2). We demonstrated this advantage by comparing the classification error of spaced $(64,8)$-mer generated by LDPC and random LSH, respectively. We observed that the classification error of Opal(64,8) is consistently lower than that of LSH(64,8). For example, when $L=200$ and mutation/sequecing error is 15\%, spaced $(64,8)$-mer generated by random LSH has a classification error of 8.31\%, while Opal$(64,8)$ method benefits from the equal coverage and gives an classification error of 7.81\%. In addition, the hierarchical $(64,32,8)$-LDPC further reduces the classification error to 7.07\%, due a better structured manner of generating LSHs for very long $k$-mers. All these comparisons are statistically significant by paired t-test.

These results demonstrate that Opal, based on the Gallager LSH method, is able to capture very long-range compositional dependency for long metagenomic fragments and shows substantial improvements over traditional $k$-mer profile methods. With the efficient design of the Gallager LDPC code which was designed for error correction in noisy communication, we observe a similar effect in metagenomic binning, that is high robustness to the errors in the data. These observations indicate the practical applicability of Opal to large-scale and noisy metagenomic sequencing datasets from environmental samples.

\subsection{Comparison to the alignment-based method}
We also compared Opal with BWA-MEM, a fully optimized read mapper.  For $L=200$, we observed that Opal works remarkably better than BWA-MEM, especially when the mutation/sequencing error rate is high. For $L=400$, the comparison is similar but the gap between Opal and BWA-MEM is smaller because the alignment-based method would benefit from the longer-range dependency. We have also compared BWA-MEM and Opal on smaller datasets with fewer species. Their performance are comparable as there is less ambiguity for BWA-MEM.

\begin{figure}[h]
  \centering
                \includegraphics[width=0.8\textwidth]{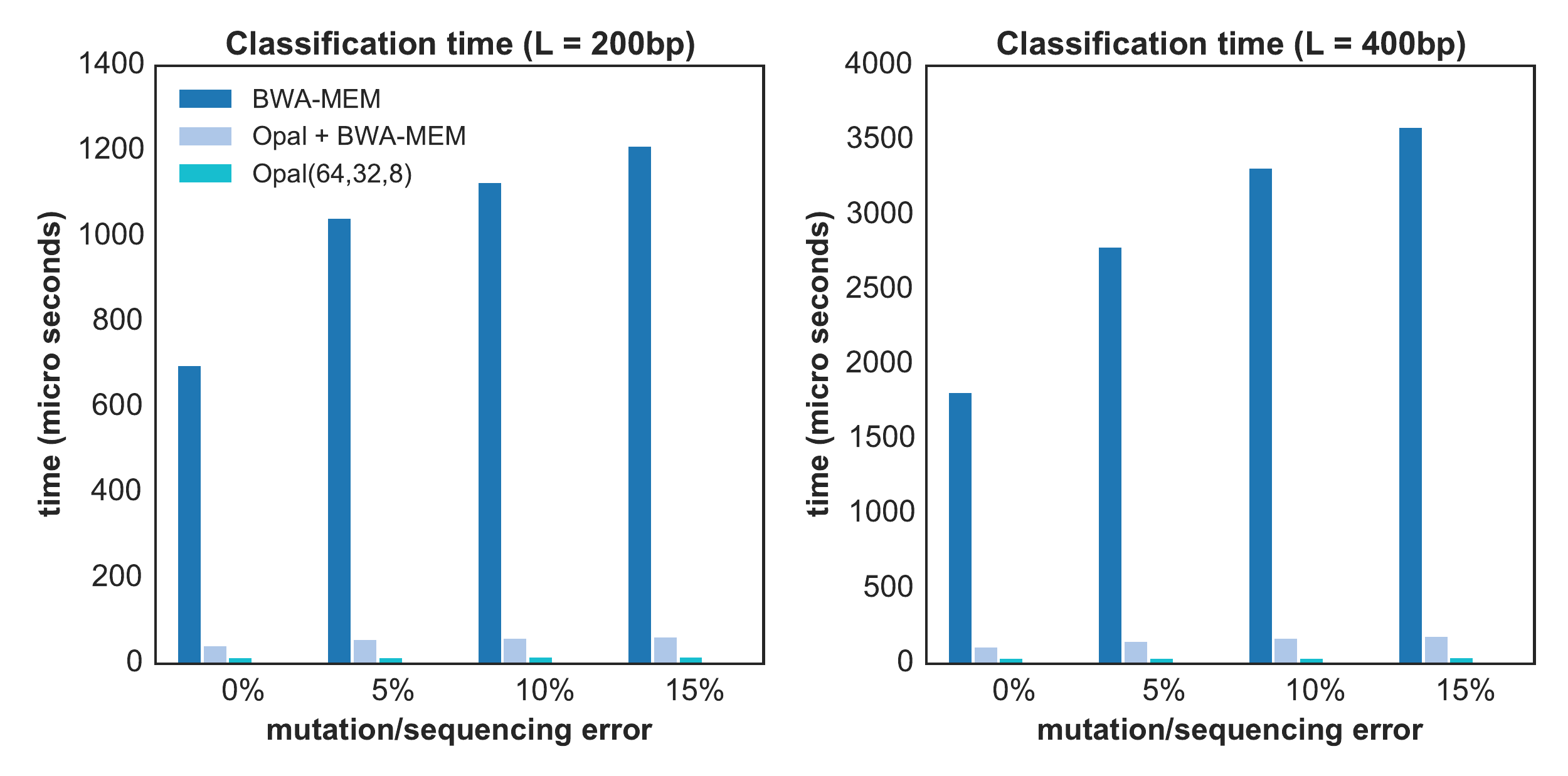}
 \caption{\textbf{Comparison of classification speed.} These figures show the time (in microseconds) required to classify each fragment using BWA-MEM, Opal, and Opal+BWA-MEM.}
  \label{fig:speed}
\end{figure}

For scalability, we compared the classification speed of our method and BWA-MEM which is highly optimized for mapping. The experiments are conducted on a machine equipped with an Intel Xeon E5-2650 CPU with 8 cores running at 2.00GHz and 32G of RAM. The classification speed of the Opal shows no variation across different mutation/sequecing error rates. The mean classification time per fragment of Opal is 13.1 and 33.8 microseconds for fragment length $L=200$ and $L=400$, respectively. However, the classification speed of BWA-MEM increases with the mutation rates. For example, BWA needs 697 microseconds per fragment at mutation/sequecing error rate 0\% and 1211 microseconds per fragment at mutation/sequecing error rate 15\%, for fragment length $L=200$. We observed that Opal performs up to near 100 times faster than BWA-MEM. In addition, since Opal only stores the weights for the classifier, its memory usage is only about $33\%$ of BWA-MEM's usage. We believe that these scalability improvements would be more significant on datasets with a more complete set of reference species and longer fragments.

\subsection{Compositional-based binning as ``coarse search"}
While our compositional-based method has numerous advantages over alignment-based methods, it is not an aligner so that it cannot provide the high-resolution mapping results for users,  which could be critical for certain downstream applications, such as comparative analysis. Given that the binning accuracy of Opal is outstanding, we can combine it with the alignment-based method in the compressive genomics manner by first applying it as a ``coarse search" step to identify a very small group of putative taxonomic origins that are ranked very top by Opal and then performing the alignment-based method as a ``fine search" to find the mapping locations and hopefully further improve the ranking predicted by Opal. In the experiment, we only picked the top 2 predicted species by Opal and see whether we can improve by using BWA-MEM subsequently. Remarkably, we found that BWA-MEM is able to further improve Opal by performing sequence alignment against the top 2 ranked reference genomes. The accuracy is very nearly optimal given the top 2 accuracy by Opal, suggesting that BWA-MEM picks the best possible taxonomic origin from the two putative species. Moreover, since we greatly reduced the search space for BWA-MEM, this integrated approach is almost 20 times faster than original BWA-MEM and also with substantially improved binning performance on noisy data. Moreover it also provides high-resolution mapping results that the compositional-based methods cannot generate. These results indicate that this natural combination of compositional-based and alignment-based methods provides metagenomic binning with high scalability and high accuracy along with high-resolution alignments.

\begin{table}[H]
  \centering
  \begin{tabular}{lcccc}
     \hline
     mutation    &    0\%  & 5\%     & 10\%   & 15\%   \\\hline
     Opal top 1        &  0.0399 & 0.0497  & 0.0542 & 0.0707 \\
     Opal top 2        &  0.0305 & 0.0417  & 0.0445 & 0.0621 \\
  Opal+BWA-MEM   &  0.0325 & 0.0426  & 0.0466 & 0.0645 \\
     \hline
   \end{tabular}
 \caption{Compositional-based binning as ``coarse search" for alignment-based method further improves binning accuracy.}
 \end{table}

\section{Discussion}
We have presented Opal, a novel compositional-based method for metagenomic binning. By drawing ideas from the Gallager LDPC code from coding theory, we designed a family of efficient and discriminative LSH functions to construct compositional features that capture the long-range dependencies within metagenomic fragments. Our method can also be seen as a dimensionality reduction approach for genomic sequence data, which extends the previously-used $k$-mer profile.

By comparing the Gallager LSH method with the traditional $k$-mer based binning method, we have demonstrated substantial improvement on large metagenomic datasets with high sequencing errors or mutations presented, which is mainly because of the theoretical properties of LSH and the efficient design of Gallager code. Clearly, when $\Sigma$ only contains $0$ and $1$ and the LSH functions maps the binary strings to vectors of $GF(2)$, our Gallager LSH method degenerates to the original LDPC code. Compared to BWA-MEM, a state-of-the-art alignment-based method, Opal achieves comparable performance when the fragments have a small number of sequencing errors or mutations and performs much better and more robust in the presence of high sequencing errors or mutations. Moreover, our method is up to two orders of magnitude faster than the alignment-based method, indicating its practical advantage on very large metagenomic datasets. Finally, we also have demonstrated that by using Opal as a ``coarse search" step to identify a small candidate set of taxonomic origins for a subsequent alignment-based method, we are able to provide metagenomic binning with high scalability and high accuracy along with high-resolution alignments. Overall, Opal enables us to perform accurate metagenomic analysis for very large metagenomic studies with greatly reduced computational cost.

In the future, we plan to further explore the improvements in metagenomic binning with Opal. For example, here we only used the simplest linear multiclass SVM, which is agnostic to the structure of taxonomy and can only provide predictions on the species level. We believe that with a structured SVM or other hierarchical classification algorithms, we will be able to perform binning on different taxonomic levels phylogenetically and even provide insights for new species or clades \cite{lmat,kraken,phymm,patil2012phylopythias}. In addition, we will compare our method to the recent indexing or nearest search-based methods, such as \cite{lmat,kraken,phymm} on larger real-world datasets and expect to see further application of the Gallager LSH method for fast similarity search-based binning. Finally, we hope to find a better way to integrate the compositional-based and alignment-based binning methods. For example, in a ``compressive genomics" manner, we can further devise the compositional-based method to handle compressed metagenomic fragments with high sequence similarity, as the LSH functions are theoretically capable for efficient coding for similar sequences. Also we hope to investigate principled guidances on how to use Opal as the ``coarse search" to better suit the subsequent alignment-based method.

\bibliographystyle{unsrt}
\bibliography{RECOMB2016}
\newpage
\section*{Supplementary Information}
\setcounter{figure}{0}
\renewcommand\thefigure{S\arabic{figure}}
\setcounter{table}{0}
\renewcommand\thetable{S\arabic{table}}
\setcounter{algorithm}{0}
\renewcommand\thealgorithm{S\arabic{algorithm}}
\begin{table}[h]
  \centering
  \begin{tabular}{ll}
     \hline
     Acetobacter pasteurianus              & Listeria monocytogenes \\
     Acinetobacter baumannii               & Methylobacterium extorquens \\
     Bacillus amyloliquefaciens            & Mycobacterium tuberculosis \\
     Bacillus anthracis                    & Mycoplasma fermentans \\
     Bacillus subtilis                     & Mycoplasma genitalium \\
     Bacillus thuringiensis                & Mycoplasma mycoides \\
     Bifidobacterium bifidum               & Mycoplasma pneumoniae \\
     Bifidobacterium longum                & Neisseria gonorrhoeae \\
     Borrelia burgdorferi                  & Propionibacterium acnes \\
     Brucella abortus                      & Pseudomonas aeruginosa \\
     Brucella melitensis                   & Pseudomonas stutzeri \\
     Buchnera aphidicola                   & Ralstonia solanacearum \\
     Burkholderia mallei                   & Rickettsia rickettsii \\
     Burkholderia pseudomallei             & Shigella flexneri \\
     Campylobacter jejuni                  & Staphylococcus aureus \\
     Corynebacterium pseudotuberculosis    & Streptococcus agalactiae \\
     Corynebacterium ulcerans              & Streptococcus equi \\
     Coxiella burnetii                     & Streptococcus mutans \\
     Desulfovibrio vulgaris                & Streptococcus pneumoniae \\
     Enterobacter cloacae                  & Streptococcus thermophilus \\
     Escherichia coli                      & Thermus thermophilus \\
     Francisella tularensis                & Treponema pallidum \\
     Helicobacter pylori                   & Yersinia enterocolitica \\
     Legionella pneumophila                & Yersinia pestis \\
     Leptospira interrogans                & Yersinia pseudotuberculosis \\
     \hline
   \end{tabular}
  \caption{List of names of the 50 microbial species used in this work.}
\end{table}

\begin{algorithm}
    \caption{Gallager's LDPC Matrix}
    \begin{algorithmic}[1]
        \State \textbf{Input:} $k$, $t$, $m$
        \State $Q \leftarrow $ all zero $(m/w)\times k$ matrix
        \For{$i \leftarrow 1$ to $m/w$}
            \For{$j \leftarrow (i-1)\times t +1$ to $i\times t$}
                \State $Q[i,j]\leftarrow1$
            \EndFor
        \EndFor
        \State choose $w-1$ uniform random $n\times n$ permutation matrix $P_i$, for $i=1,\ldots,w-1$.
        \State $H=[Q; QP_1; \ldots; QP_{w-1}]^T$
        \State \textbf{Output:} Gallager's LDPC Matrix $H$
    \end{algorithmic}
    \label{alg:gallager}
\end{algorithm}

\begin{algorithm}
    \caption{Removing 4-Cycles}
    \begin{algorithmic}[1]
        \State \textbf{Input:} Gallager's LDPC Matrix $H$
        \Repeat
            \For{$i \leftarrow 1$ to $k-1$}
                \For{$j \leftarrow i+1$ to $k$}
                    \If{$|H[:,i]\cup H[:,j]| \ge 2$}\Comment{check if 4-cycle exists}
                        \State $ridx \leftarrow $ row index of the first same element in $H[:,i]$ and $H[:,j]$.
                        \State $b \leftarrow \lceil ridx / (m/w)\rceil$
                        \State swap the elements of $H[:,i]$ and $H[:,j]$ that belong to the $b$-th block.
                    \EndIf
                \EndFor
            \EndFor
        \Until{no 4-cycle}
        \State \textbf{Output:} 4-Cycle-free Gallager's LDPC Matrix $H$
    \end{algorithmic}
    \label{alg:cycle}
\end{algorithm}

\clearpage

\begin{figure}[h]
  \centering
  \begin{subfigure}{0.49\textwidth}
                \includegraphics[width=\textwidth]{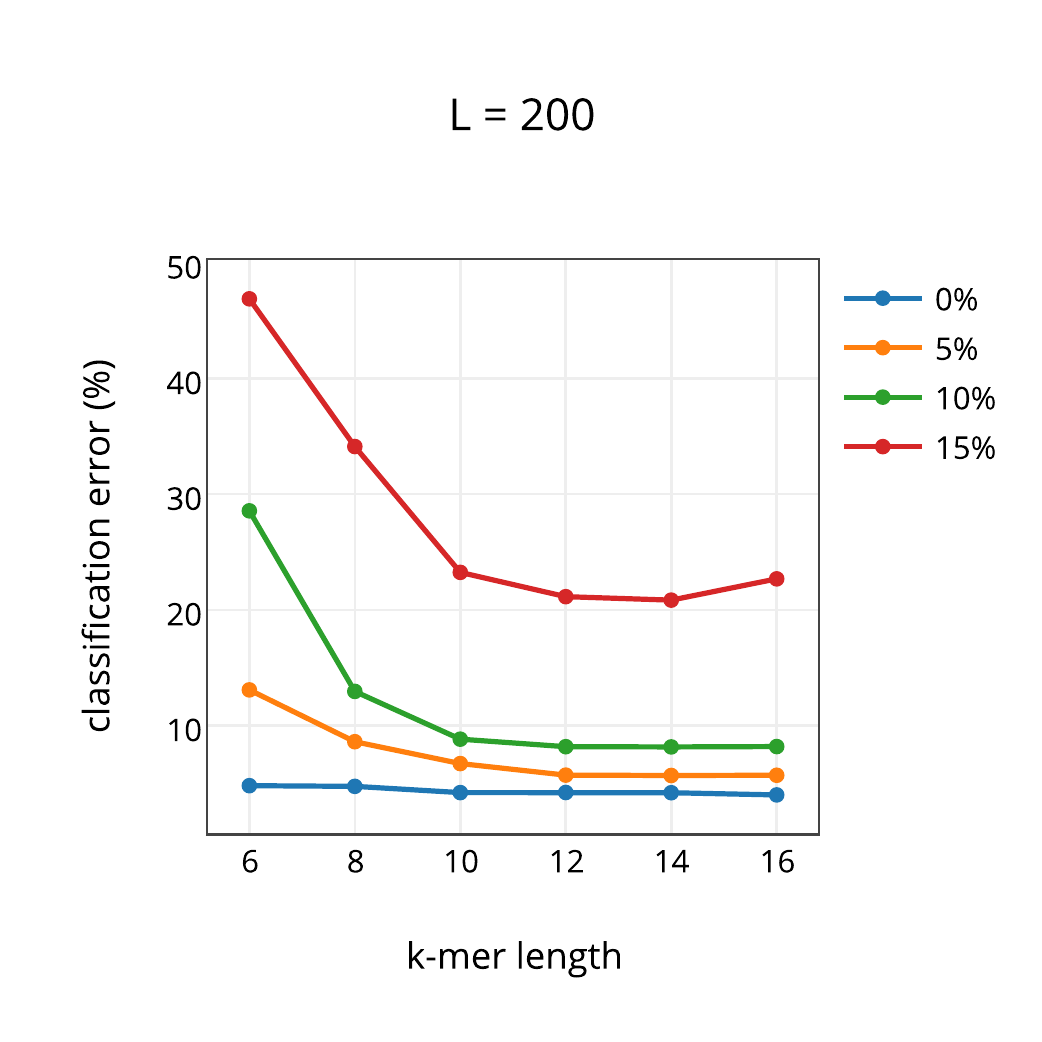}
  \end{subfigure}
\begin{subfigure}{0.49\textwidth}
                \includegraphics[width=\textwidth]{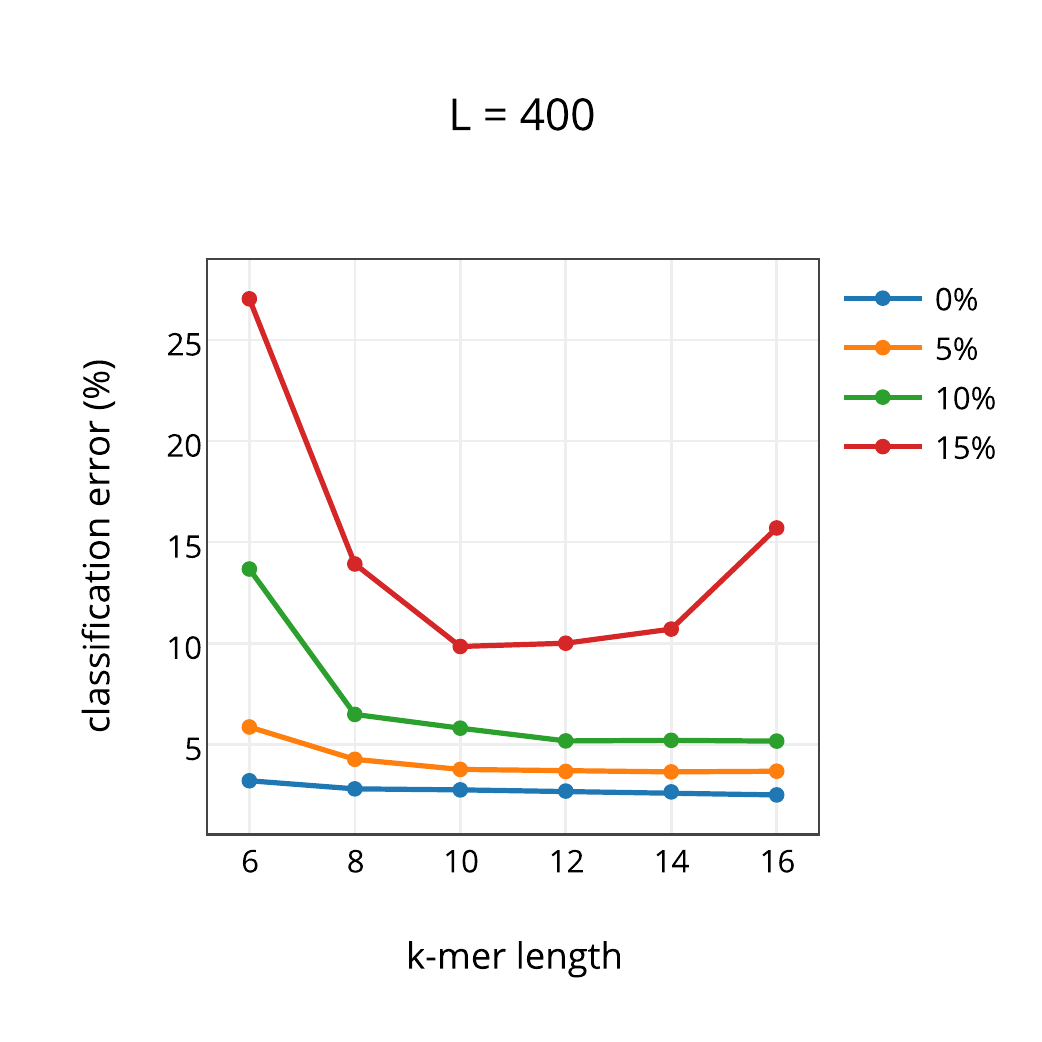}
        \end{subfigure}
  \caption{Performance of $k$-mer profiles with different $k$ values on the 50-species dataset with $c=0.1$. }
  \label{fig:kmerLen}
\end{figure}

\begin{figure}[h]
  \centering
  \begin{subfigure}{0.49\textwidth}
                \includegraphics[width=\textwidth]{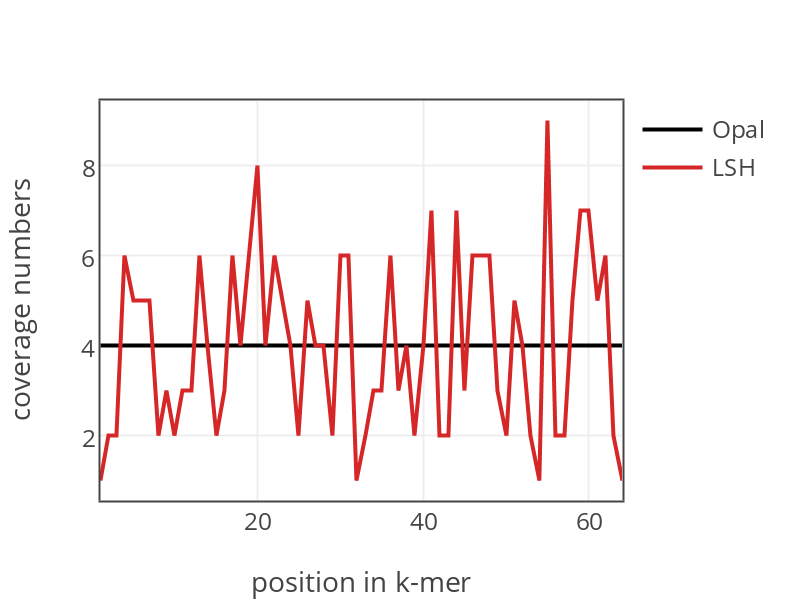}
  \end{subfigure}
\begin{subfigure}{0.49\textwidth}
                \includegraphics[width=\textwidth]{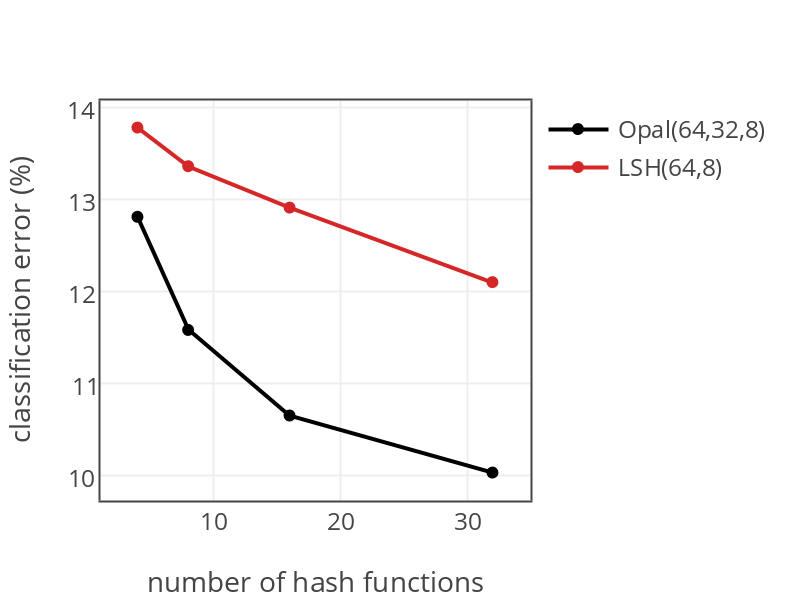}
        \end{subfigure}
  \caption{Comparison of position coverage between LSH and Gallager LSH. The left figure shows how many times a position being covered when spaced $(64,8)$-mers are generated by $m$ uniformly sampled LSH and Gallager LSH functions, respectively. The right figure compares the relationship between the number of hash functions and the classification performance. The experiment is done on a 20-species dataset.}
  \label{fig:cover}
\end{figure}

\end{document}